# Temperature Dependent Characteristics and Electrostatic Threshold Voltage Tuning of Accumulated Body MOSFETs

ABM Hasan Talukder, *Student Member, IEEE*, Brittany Smith, Mustafa Akbulut, Faruk Dirisaglik, Helena Silva, *Senior Member, IEEE,* and Ali Gokirmak, *Senior Member, IEEE*

***Abstract*—Narrow-channel accumulated body nMOSFET devices with p-type side-gates surrounding the active area have been electrically characterized between 100 and 400 K with varied side-gate biasing ($V_{side}$). The subthreshold slope (SS) and drain induced barrier lowering (DIBL) decrease and threshold voltage ($V_t$) increases linearly with reduced temperature and reduced side-gate bias. Detailed analysis on a 27 nm x 78 nm (width x length) device show SS decreasing from 115 mV/dec at 400 K to 90 mV/dec at 300 K and down to 36 mV/dec at 100 K, DIBL decreasing by approximately 10 mV/V for each 100 K reduction in operating temperature, and $V_t$ increasing from 0.42 V to 0.61 V as the temperature is reduced from 400 K to 100 K. $V_t$ can be adjusted from ~ 0.3 V to ~ 1.1 V with ~0.3 V/V sensitivity by depletion or accumulation of the body of the device using $V_{side}$. This high level of tunability allows electronic control of $V_t$ and drive current for variable temperature operation in a wide temperature range with extremely low leakage currents (< $10^{-13}$ A).**

## I. Introduction

ELECTROSTATIC control of the potential barrier between the source and the drain of metal oxide field effect transistors (MOSFETs), and the emitter and collector of bipolar junction transistors (BJTs) determine the minimum feature sizes that can be successfully implemented for a given temperature of operation. Since the charges controlling the barrier height in MOSFETs are electrically isolated from the channel, there is no need for continuous charge injection from the control terminal, unlike BJTs. Ultra-thin-body silicon on insulator (SOI) [1]–[5], FinFET [6]–[9], tri-gate [10]–[15], gate-all-around [16]–[23] and multi-gate [24], [25] structures provide significant improvements in electrostatic control of the source-barrier in MOSFETs and suppress or eliminate the interface leakage currents [26]. Thin-body SOI and gate-all-around devices suffer from floating body effects [27], [28], where majority carriers trapped in the active region reduce the source-barrier (similar to charging of the base in BJTs) and cause soft errors. The active regions of bulk MOSFETs have good electrical and thermal contact with the substrate, allowing for efficient charge and heat removal. Electrostatic control of the source-barrier using substrate/body [23], [29]–[36] or back-gate biasing [37]–[39] allows for dynamic control of threshold voltage. In the case of narrow channel devices, an independently controlled side-gate structure that surrounds the body of a bulk MOSFET can be used to accumulate the body of the device and increase electrical coupling of the body to the channel, increase the source barrier and suppress short-channel effects (Fig. 1).

Fabrication of these devices were performed by M. Akbulut and F. Dirisaglik at IBM T J Watson Research Center supported by U.S. National Science Foundation (NSF) ECCS Award 0824171. The electrical characterization and analysis were performed by A. Talukder and B. Smith supported by the NSF ECCS Award 1711626. H. Silva and A. Gokirmak supervised the project and contributed to design of experiments, analysis, and manuscript preparation.
Authors are with the Electrical and Computer Engineering Department, University of Connecticut, Storrs, CT 06269 USA (e-mail: talukder@uconn.edu; ali.gokirmak@uconn.edu).

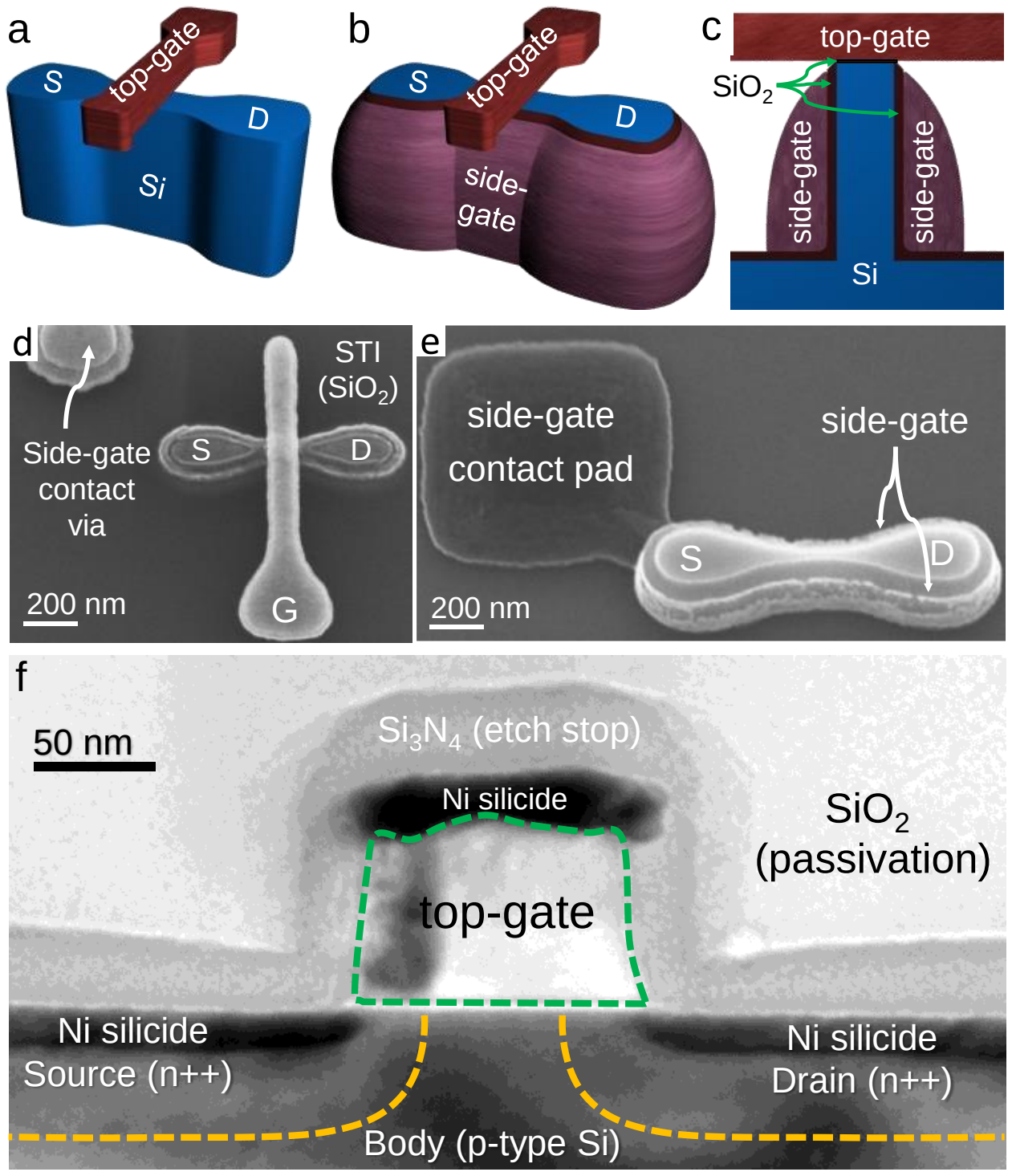

Fig. 1. Schematic drawings of a conventional narrow-channel bulk-Si MOSFET (a), accumulated body MOSFET with side-gate structure (b) and its cross-section view (c). SEM image of the top view of the accumulated-body MOSFET showing the side-gate contact via after shallow trench isolation (STI) and gate formation (d) and the perspective view of the MOSFET showing the side-gate contact pad before STI (e), TEM image of a cross-sectional cut in the Source-Drain direction [24] (f).

We had first reported a side-gated MOSFET structure with silicon nitride ($Si_3N_4$ / $Si_{3+x}N_4$) side-gate dielectric and field isolation (compatible with removal of sacrificial $SiO_2$ using HF to open tunnels and release structures) as a very-low leakage device that can be monolithically integrated with micro-/nano-fluidics for high-sensitivity sensors for sequencing of biomolecules [40], [41]. Noticing the success of these devices in suppressing interface leakage currents (below 1 fA), improvements in drain induced barrier lowering (DIBL) and subthreshold slope (SS), and extreme control of threshold voltage ($V_t$), we fabricated higher-performance versions of these devices with p+ doped side-gates, $SiO_2$ side-gate dielectrics, and $SiO_2$ field isolations for accumulated body operation. These accumulated-body devices show dramatic $V_t$ tuning and improved performance characteristics [42]–[45].

Transistors are thermal devices; the barrier height, the energy distribution of the carriers in the source and drain reservoirs and lifetime of trapped charges [46] depend on temperature. Carrier mobilities also increase with reduced temperature [23], [24], [47]–[49]. Hence, transistors can be operated with lower power and at a higher speed under lower temperatures. While high temperature operation is necessary for certain applications [50]–[53], cryogenic operation of high-performance VLSI circuits has been considered for a long time [23], [24], [47]–[49], [54]–[58]. Recent advances and growing interest in quantum computing have also led to renewed interest in cryogenic electronics for peripheral circuitry. However, significant changes in the operation temperature of the whole or part of the circuit introduces additional complexity in VLSI design and in computer architecture [59]–[61]. In this work, we characterized accumulated body devices in the 100 K to 400 K range, observed the performance characteristics, and explored the ability to dynamically tune $V_t$ as the temperature is changed and the temperature range these devices can reliably operate at.

## II. Device Structure and Fabrication

The narrow-channel accumulated body nMOSFETs were fabricated using a conventional bulk silicon process, with additional steps to form the p+ doped side-gate structure and its contact. The side-gate structure is formed around the body using a side-wall process [44], [62]. The side-gate contact is formed in the same etch step by using a lithographically defined region slightly offset from the active region. The side-gates are isolated from the body by 9 nm thermally grown $SiO_2$ and the top-gate is isolated from the active area by 3.9 nm thermally grown $SiO_2$. The body is doped with $1\times10^{17}$ $cm^{-3}$ boron, the side-gate with $1\times10^{20}$ $cm^{-3}$ boron, the source and drain with $1\times10^{20}$ $cm^{-3}$ phosphorous using ion-implantation. Details of the device fabrication, room temperature electrical characterization and simulation results, including high temperature behavior, can be found in [43], [44], and [62].

## III. Electrical Characterization

All the experiments were carried out using an Agilent 4156C Parameter Analyzer between 100 K and 400 K in 25 K intervals in a Janis cryogenic probe station under vacuum, in dark. The characteristics presented here are from measurements performed on a device with width ($W$) × length ($L$) = 27 nm ×

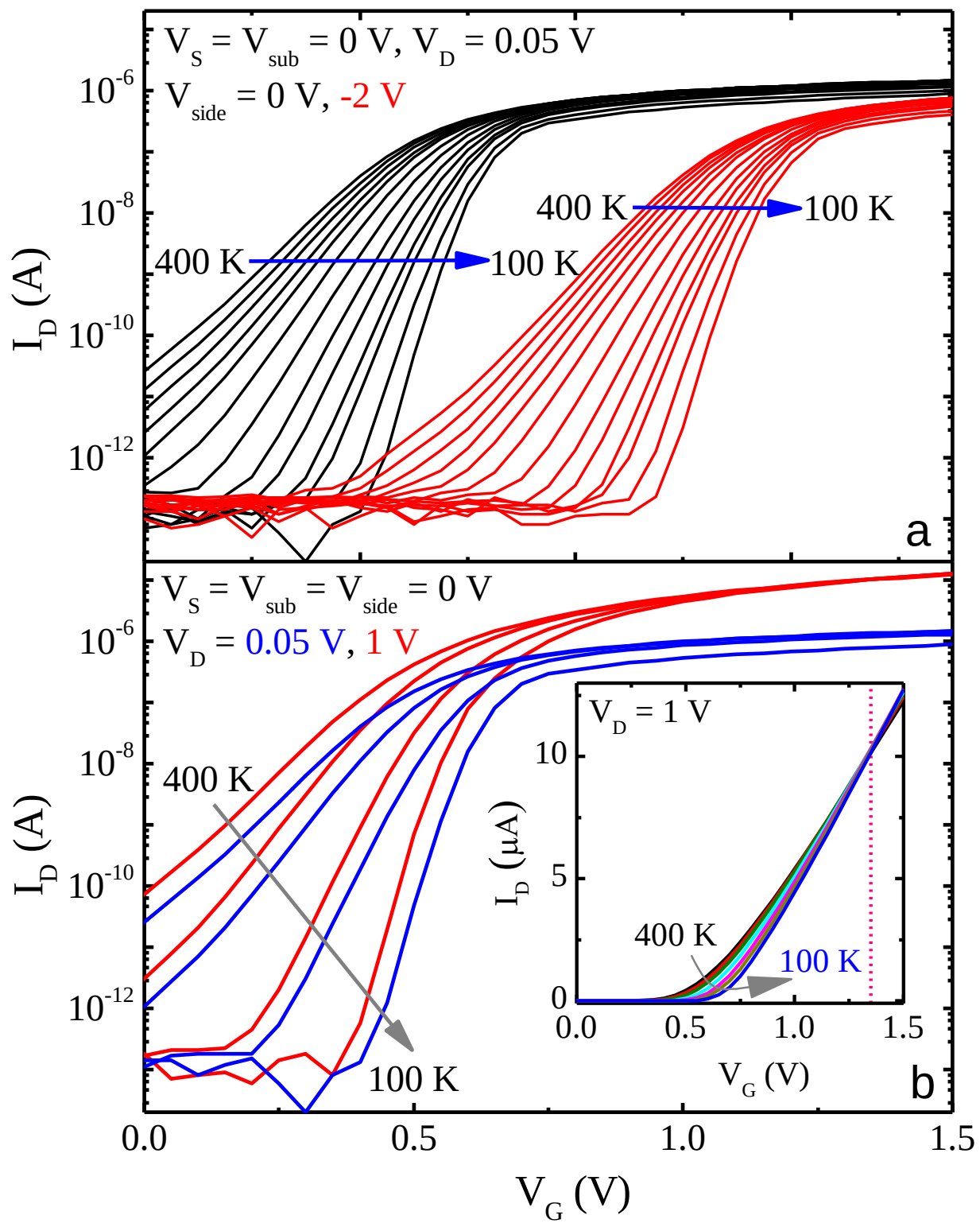

Fig. 2. Temperature and side-gate bias dependent transfer characteristics of a $W \times L$ = 27 nm × 78 nm device. The change in $V_t$ induced by the side-gate bias is much larger than the changes due to temperature (a). Temperature sensitivity of $V_t$ decreases with reduced $V_{side}$. The transfer curves converge at $V_G \approx 1.35$ V for high drain bias (inset).

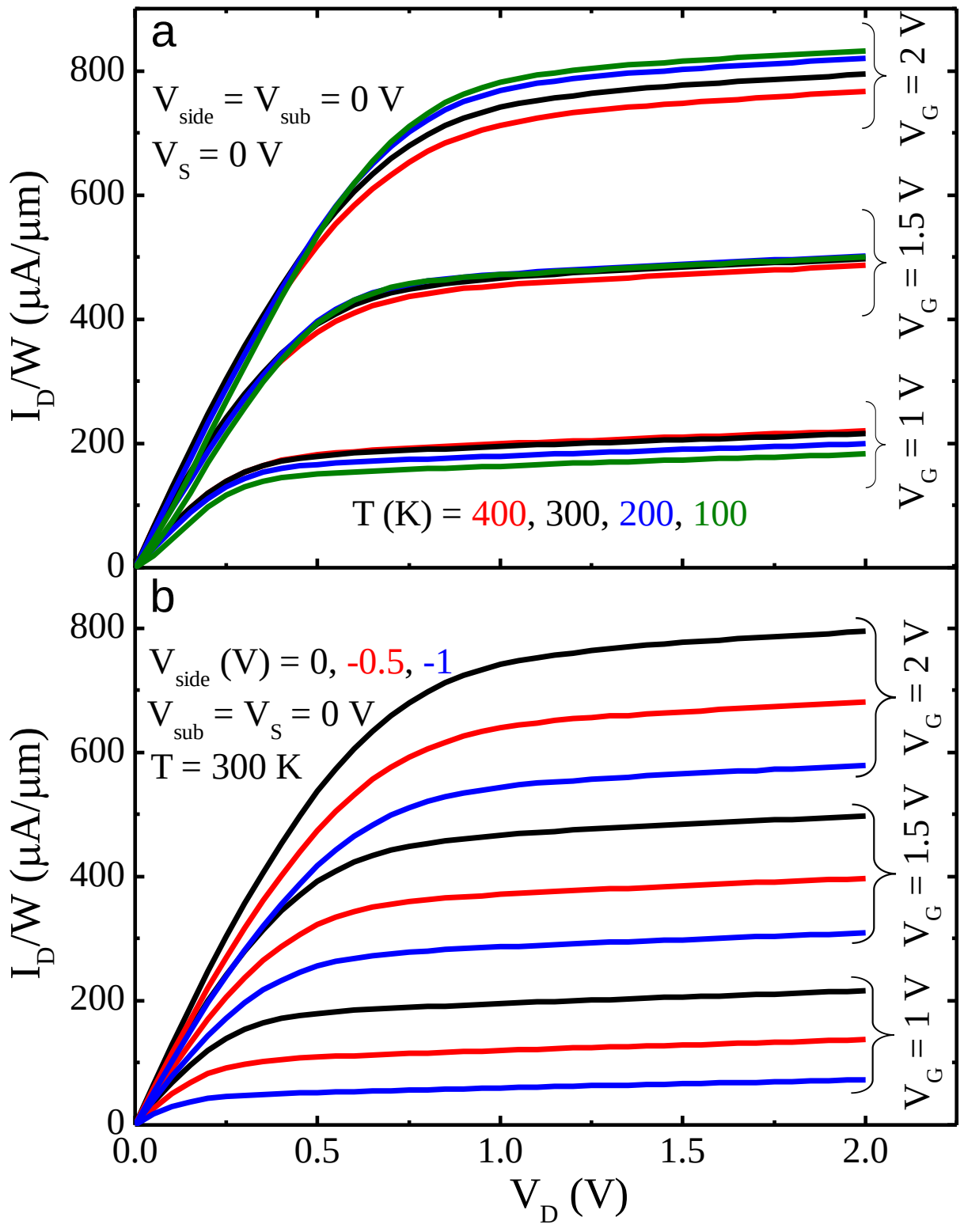

Fig. 3. Temperature (a) and side-gate bias (b) dependent output characteristics of a $W \times L$ = 27 nm × 78 nm device. Saturation currents are insensitive to temperature for $V_G \approx 1.5$ V (a). Negative side-gate bias causes the saturation current to decrease (b).

78 nm, with grounded source and substrate contacts ($V_S$ = $V_{sub}$ = 0 V) unless stated otherwise (Fig. 2 and Fig. 3).

The body/side-oxide interfaces of these devices are in accumulation, even with $V_{side}$ = 0 V since the side gates are p+ doped (unlike [40], [41]), thus suppressing recombination at the interface defects, blocking interface leakage currents, and reducing the depletion depth under the active region. Negative bias on the side-gate increases the hole concentration at the side-interfaces, shrinks the depletion depth further, increasing the electrostatic control of the body and suppressing control of the drain potential on the source barrier [41], [44], hence, improving DIBL and SS and increasing $V_t$.

The sensitivity of $V_t$ to temperature ($\Delta V_t/\Delta T$) is significantly reduced for negative side-gate bias while sensitivity of $V_t$ to $V_{side}$ ($\Delta V_t/\Delta V_{side}$) decreases slightly with reduced temperature as shown in Fig. 4(a) and (b). Subthreshold slope (SS), drain induced barrier lowering (DIBL= $\Delta V_t/\Delta V_D$), and maximum transconductance ($g_m$) show exponential responses to $V_{side}$ and a linear response to temperature [Fig. 4(c)-(h)]. Negative side-gate bias increases $V_t$ linearly [Fig. 4(b)], significantly more than the change caused by temperature in 100 K to 400 K range [Fig. 4(a)]. The observed reduction in drive current [Fig. 4(i)-(l)] appears to be mostly due to increase in $V_t$ for high drain biases. The saturation current ($I_{D\text{-sat}}$) is directly proportional to temperature at lower gate voltages and inversely proportional to temperature at higher gate voltages, and insensitive to temperature for $V_G \approx 1.35$ V [Fig. 2 (inset) and Fig. 4(i)]. This is due to the interplay of increased carrier injection over the source-barrier and reduced mobility with increasing temperature [63]. The response of $I_{D\text{-sat}}$ to $V_{side}$ is monotonous [Fig. 4(j)].

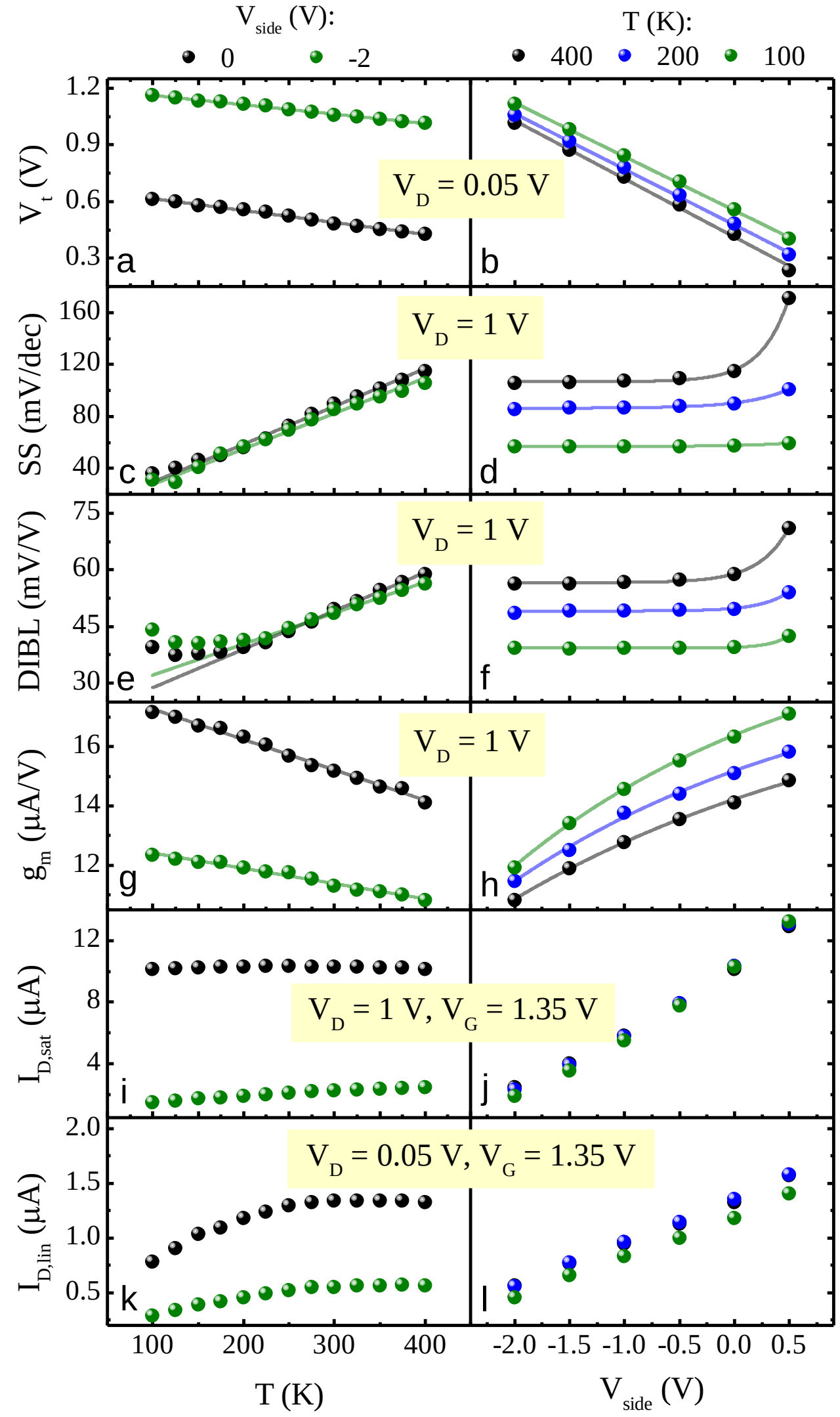

Fig. 4. Temperature and side-gate biasing response of threshold voltage ($V_t$), subthreshold slope (SS) and drain induced barrier lowering (DIBL), maximum transconductance ($g_m$), current drive in saturation and linear regimes extracted from transfer characteristics presented in Fig. 2. $V_t$ values are determined from the linear extrapolation of $I_D$ at the maximum transconductance point. SS and DIBL are calculated at $V_G$ corresponding to the steepest SS. SS improves by ~ 29 mV/dec for each 100 K.

## IV. Threshold Voltage Tuning

Accumulated body MOSFET offers an independent control through the side-gate which can be used to achieve a certain threshold voltage at different operation temperatures or to compensate for threshold drifts in unwanted local hotspots. The approximately linear response of $V_t$ to $V_{side}$ and $T$ (Fig. 4) can be used to predict $V_{side}$ values required to maintain a constant $V_t$ in a wide temperature range (Fig. 5).

The $V_t$ sensitivity to side-gate bias varies only slightly as a function of temperature, from ~ -308 mV/V at 400 K to ~ -278 mV/V at 100 K, between -2 and +0.5 V side-gate bias. This side biasing sensitivity of ~ 300 mV/V allows for threshold voltage tuning across a wide range of voltages. Linear fits to the set of $V_t$ vs $V_{side}$ curves in the 100 – 400 K range allow us to determine the required $V_{side}$ value to tune $V_t$ at a given temperature (Fig. 6). We then repeat this procedure to estimate $V_{side}$ vs $T$ relationships for any desired $V_t$ values (Fig. 7). To validate our procedure, we repeated the $I_D$-$V_G$ measurements between 100- and 400 K while applying the calculated $V_{side}$ for each temperature to tune $V_t$ to example target values of 0.5 V, 0.7 V, and 0.9 V. Fig. 8 shows the resulting $V_t$ to be very close to the target values. Each *I-V* was performed once with the calculated $V_{side}$, after the previous one, with no need for any adjusting iterations, indicating minimal charge trapping on the side-gate dielectric, $SiO_2$ and $Si/SiO_2$ interfaces.

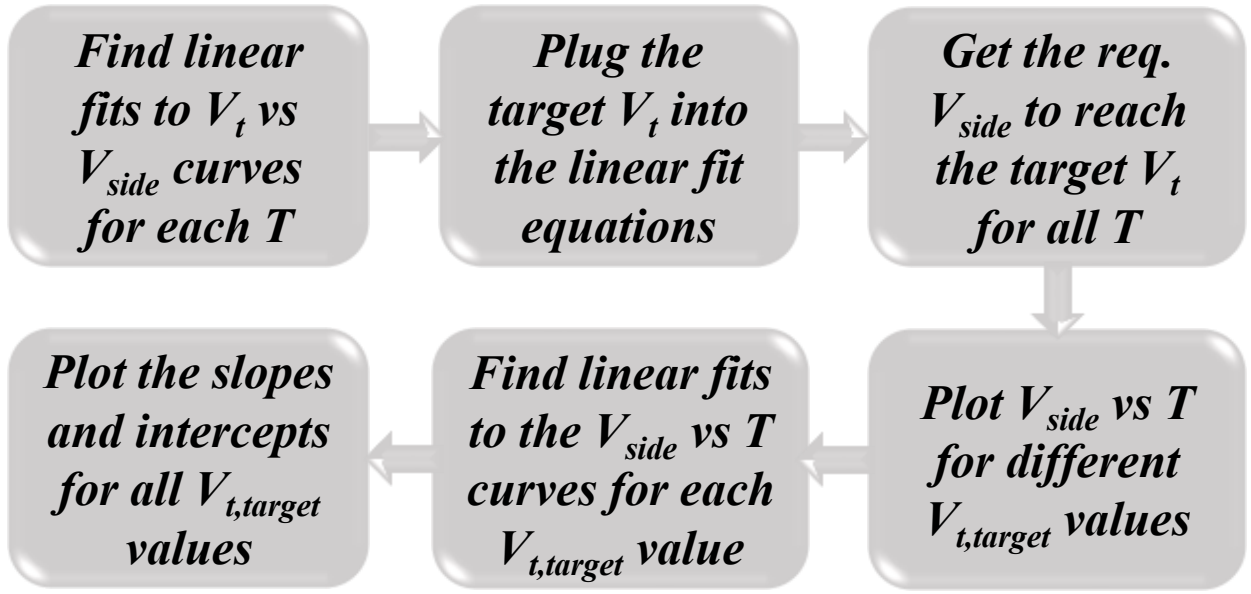

Fig. 5. A flow diagram of the procedure used for $V_t$ tuning process.

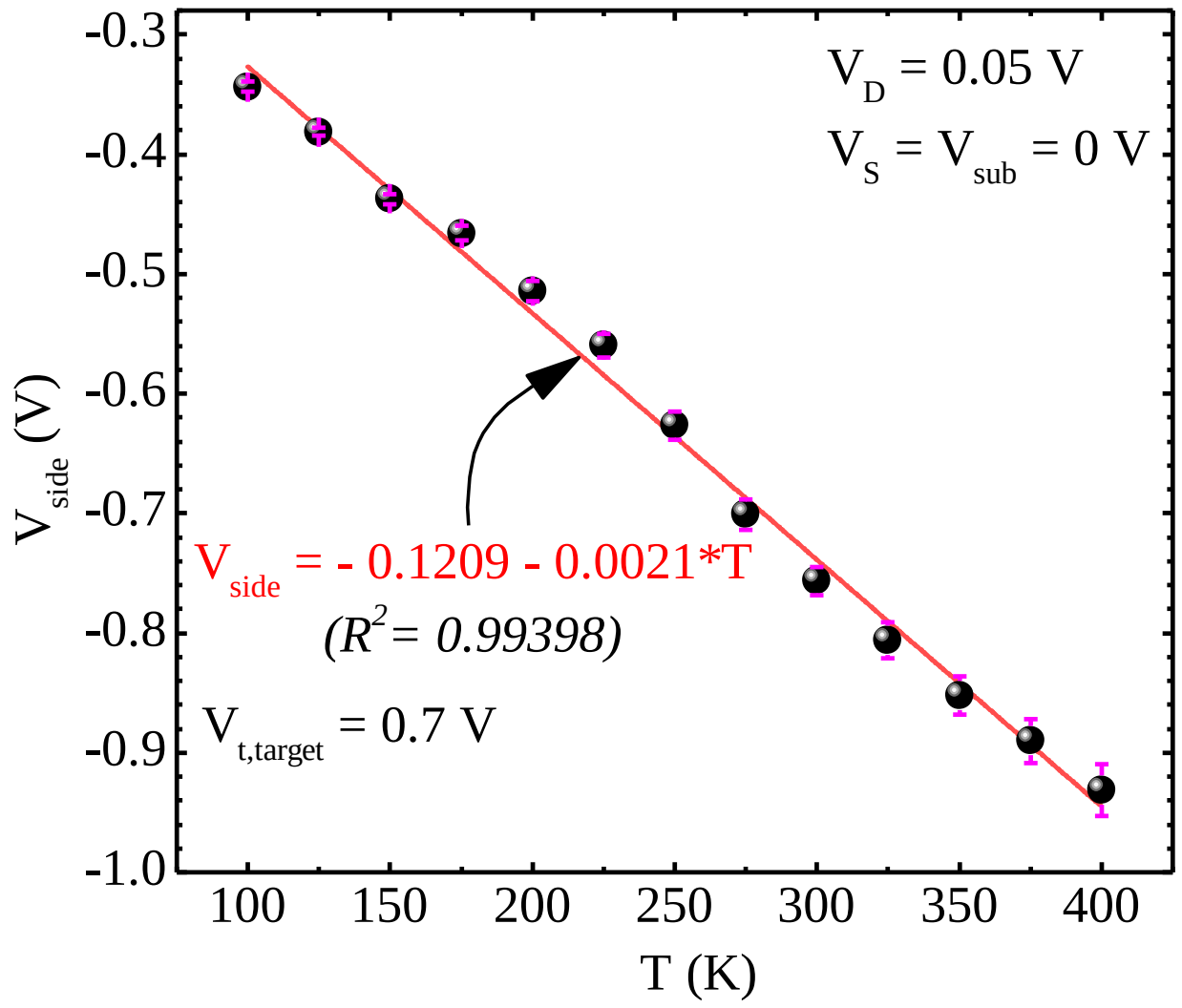

Fig. 6. The side-gate biases required to keep the threshold voltage fixed to 0.7 V across the entire temperature range (100-400 K). The linear fit shows a required rate of 2.1 mV/K $V_{side}$ change to tune the threshold voltage to 0.7 V with a drain bias of 50 mV and the source and the substrate grounded for the device $W \times L$ = 27 nm × 78 nm.

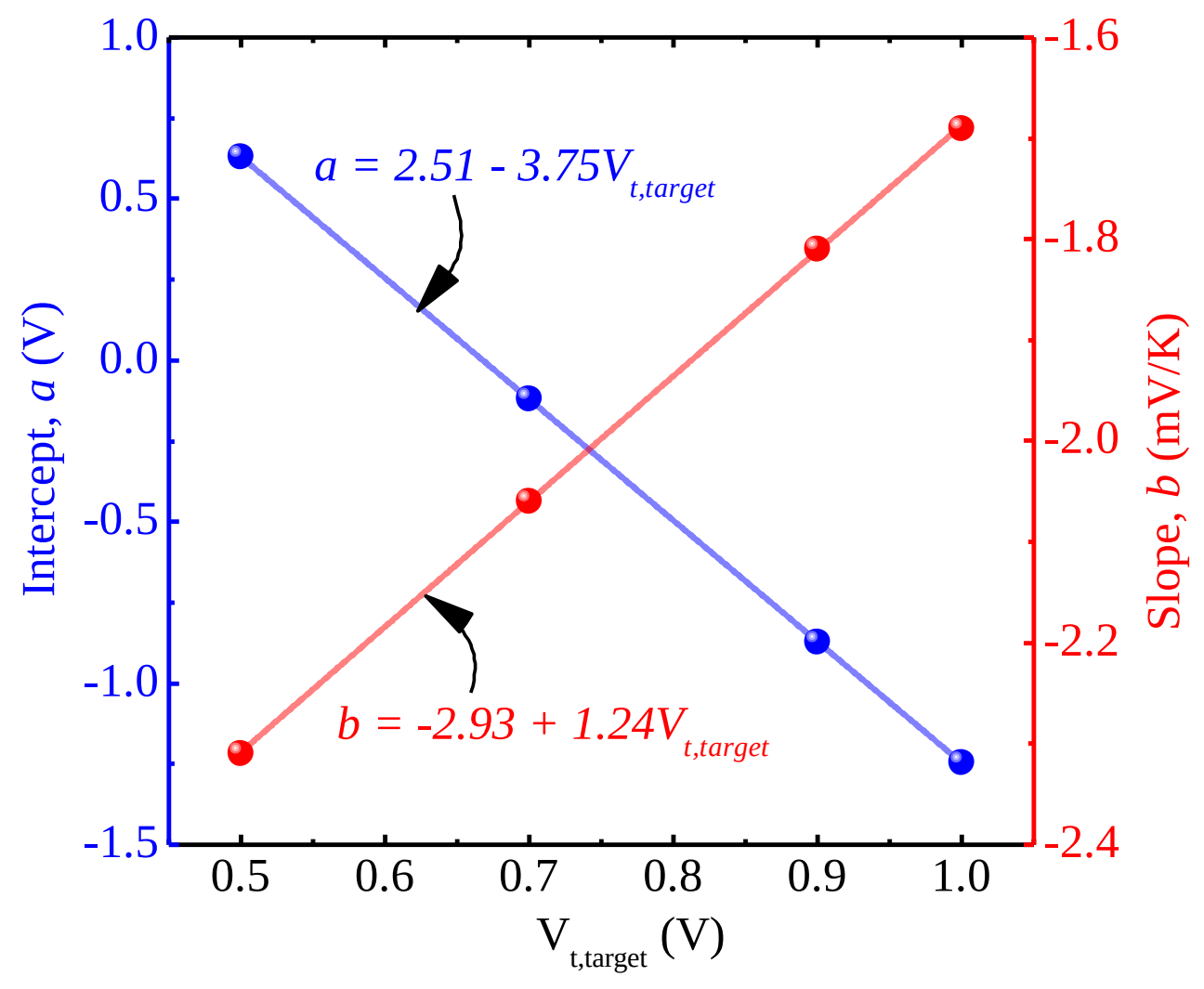

Fig. 7. Fitting parameters (slope and intercept) for the $V_{side}$ vs $T$ at different $V_{t,target}$. The parameters change linearly with the desired $V_t$ value. From the fit equations, we can find the values of the slope and the intercept for $V_{side}$ vs $T$ graph for the device $W \times L$ = 27 nm × 78 nm.

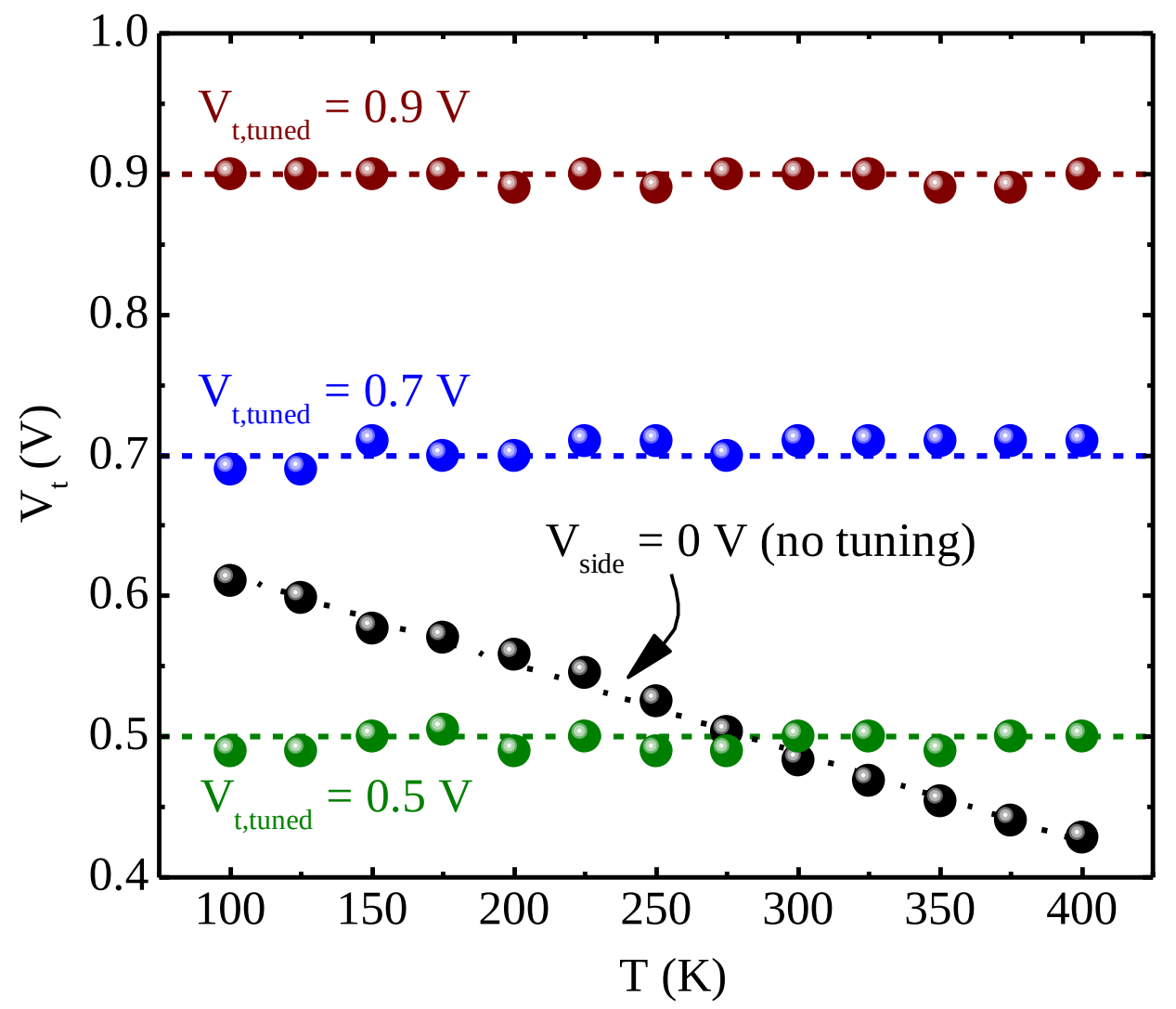

Fig. 8. The calculated threshold voltages after applying the side-gate biases required to keep $V_t$ fixed to 0.5 V, 0.7 V, and 0.9 V, separately, at each temperature point for the device $W \times L$ = 27 nm × 78 nm. The tuned $V_t$ values compared to the zero-side-gate bias points to the proper side-gate biasing compensating for a ~6 mV change in $V_t$ for a 10 K temperature reduction.

## V. Summary

Temperature dependent electrical characteristics of narrow side-gated accumulated body MOSFETs in the 100 K and 400 K range are presented. The measurements show strong and very stable temperature and side-gate bias dependence of the transistor characteristics. Subthreshold slope, drain induced barrier lowering, threshold voltage, and leakage current, all exhibit significant improvements with decreasing temperature and side-gate bias, due to mitigation of short-channel effects. Threshold voltage can be tuned with ~ -300 mV/V sensitivity by side-gate biasing, with a fairly linear response. Hence, it is relatively easy to maintain a constant threshold voltage in a wide temperature range or dramatically change the threshold voltage in part or the whole circuit by adjusting the side-gate biases. Excessively negative side-gate biases are expected to result in degradation of the current drive, due to increased surface roughness scattering, which are more pronounced at lower temperatures [64].

Since the depletion depth and the threshold voltage are controlled by the side-gate bias, an undoped body can also be used, reducing any threshold voltage variations due to random dopant fluctuations [65]. Furthermore, the depletion region can be reduced significantly more than what could be achieved with body doping or biasing and the threshold voltage can be increased above what is achievable by doping body or body biasing. The additional electrostatic control over performance parameters is expected to enable reliable circuit operation in a wide temperature range.

## References

[1] T. C. Hsiao and J. C. S. Woo, "Subthreshold Characteristics of Fully Depleted Submicrometer SOI MOSFET's," *IEEE Trans. Electron Devices*, vol. 42, no. 6, pp. 1120–1125, 1995.

[2] Y. Tian, R. Huang, X. Zhang, and Y. Wang, "A novel nanoscaled device concept: Quasi-SOI MOSFET to eliminate the potential weaknesses of UTB SOI MOSFET," *IEEE Trans. Electron Devices*, vol. 52, no. 4, pp. 561–568, 2005.

[3] A. Majumdar, Z. Ren, S. J. Koester, and W. Haensch, "Undoped-body extremely thin SOI MOSFETs with back gates," *IEEE Trans. Electron Devices*, vol. 56, no. 10, pp. 2270–2276, 2009.

[4] A. Griffoni *et al.*, "Electrical-based ESD characterization of ultrathin-body SOI MOSFETs," *IEEE Trans. Device Mater. Reliab.*, vol. 10, no. 1, pp. 130–141, 2010.

[5] M. K. Anvarifard and A. A. Orouji, "Enhanced critical electrical characteristics in a nanoscale low-voltage SOI MOSFET with dual tunnel diode," *IEEE Trans. Electron Devices*, vol. 62, no. 5, pp. 1672–1676, 2015.

[6] D. Hisamoto *et al.*, "FinFET — A Self-Aligned Double-Gate MOSFET," vol. 47, no. 12, pp. 2320–2325, 2000.

[7] D. S. Woo *et al.*, "Electrical characteristics of FinFET with vertically nonuniform source/drain doping profile," *IEEE Trans. Nanotechnol.*, vol. 1, no. 4, pp. 233–236, 2002.

[8] D. M. Fried, J. S. Duster, and K. T. Kornegay, "Improved independent gate N-type FinFET fabrication and characterization," *IEEE Electron Device Lett.*, vol. 24, no. 9, pp. 592–594, 2003.

[9] E. Yu, K. Heo, and S. Cho, "Characterization and Optimization of Inverted-T FinFET under Nanoscale Dimensions," *IEEE Trans. Electron Devices*, vol. 65, no. 8, pp. 3521–3527, 2018.

[10] B. S. Doyle *et al.*, "High performance fully-depleted tri-gate CMOS transistors," *IEEE Electron Device Lett.*, vol. 24, no. 4, pp. 263–265, 2003.

[11] J. Kavalieros *et al.*, "Tri-gate transistor architecture with high-k gate dielectrics, metal gates and strain engineering," *Dig. Tech. Pap. - Symp. VLSI Technol.*, vol. 00, no. c, pp. 50–51, 2006.

[12] X. Sun *et al.*, "Tri-gate bulk MOSFET design for CMOS scaling to the end of the roadmap," *IEEE Electron Device Lett.*, vol. 29, no. 5, pp. 491–493, 2008.

[13] M. Saitoh, K. Ota, C. Tanaka, K. Uchida, and T. Numata, "10nm-diameter tri-gate silicon nanowire MOSFETs with enhanced high-field transport and V th tunability through thin BOX," *Dig. Tech. Pap. - Symp. VLSI Technol.*, vol. 3025, no. 2006, pp. 11–12, 2012.

[14] S. H. Kim *et al.*, "High performance tri-gate extremely thin-body inas-on-insulator MOSFETs with high short channel effect immunity and $V_{th}$ tunability," *IEEE Trans. Electron Devices*, vol. 61, no. 5, pp. 1354–1360, 2014.

[15] S. H. Wu, C. H. Yu, and P. Su, "New Findings on the Drain-Induced Barrier Lowering Characteristics for Tri-Gate Germanium-on-Insulator p-MOSFETs," *IEEE J. Electron Devices Soc.*, vol. 3, no. 6, pp. 441–446, 2015.

[16] K. Nayak, S. Agarwal, M. Bajaj, P. J. Oldiges, K. V. R. M. Murali, and V. R. Rao, "Metal-gate granularity-induced threshold voltage variability and mismatch in Si gate-all-around nanowire n-MOSFETs," *IEEE Trans. Electron Devices*, vol. 61, no. 11, pp. 3892–3895, 2014.

[17] P. Zheng, D. Connelly, F. Ding, and T. J. K. Liu, "FinFET Evolution Toward Stacked-Nanowire FET for CMOS Technology Scaling," *IEEE Trans. Electron Devices*, vol. 62, no. 12, pp. 3945–3950, 2015.

[18] D. J. Moni and T. J. Vinitha Sundari, "Performance analysis of junctionless gate all around tunnel field effect transistor," *Proc. 3rd Int. Conf. Devices, Circuits Syst. ICDCS 2016*, pp. 262–266, 2016.

[19] S. Rewari, V. Nath, S. Haldar, S. S. Deswal, and R. S. Gupta, "Dual metal (DM) Insulated Shallow Extension (ISE) Gate All Around (GAA) MOSFET to reduce gate induced drain leakages (GIDL) for improved analog performance," *Proc. 2nd Int. Conf. 2017 Devices Integr. Circuit, DevIC 2017*, no. Dm, pp. 401–406, 2017.

[20] S. Jha and S. K. Choudhary, "Impact of Device Parameters on the Threshold Voltage of Double-Gate, Tri-Gate and Gate-All-Around MOSFETs," *Proc. Int. Conf. 2018 IEEE Electron Device Kolkata Conf. EDKCON 2018*, no. Ic, pp. 596–599, 2018.

[21] A. Menaria, R. Pandey, and R. Kumar, "An investigation on a triple material double gate cylindrical gate all around (TMDG-CGAA) MOSFET for enhanced device performance," *Mater. Today Proc.*, vol. 30, pp. 123–127, 2020.

[22] R. Yadav, K. Ahuja, and D. S. Rathee, "Performance Enhancement of GAA Multi-Gate Nanowire with Asymmetric Hetero-Dielectric Oxide," *Silicon*, 2021.

[23] J. Gu *et al.*, "Cryogenic transport characteristics of P-type gate-all-around silicon nanowire MOSFETs," *Nanomaterials*, vol. 11, no. 2, pp. 1–11, 2021.

[24] F. Balestra and G. Ghibaudo, "Physics and performance of nanoscale semiconductor devices at cryogenic temperatures," *Semicond. Sci. Technol.*, vol. 32, no. 2, 2017.

[25] F. A. Herrera, M. Miura-Mattausch, T. Iizuka, H. Kikuchihara, Y. Hirano, and H. J. Mattausch, "Modeling of Short-Channel Effect on Multi-Gate MOSFETs for Circuit Simulation," *3rd Int. Symp. Devices, Circuits Syst. ISDCS 2020 - Proc.*, pp. 20–23, 2020.

[26] M. Roy and Mahmoodi-Meimand, "Leakage current mechanisms and leakage reduction techniques in deep-submicrometer CMOS circuits," *Proc. IEEE*, vol. 91, no. 2, p. 303, 2003.

[27] P. F. Lu *et al.*, "Floating-body effects in partially depleted SOI CMOS circuits," *IEEE J. Solid-State Circuits*, vol. 32, no. 8, pp. 1241–1252, 1997.

[28] A. Wei, M. J. Sherony, and D. A. Antoniadis, "Effect of floating-body charge on SOI MOSFET design," *IEEE Trans. Electron Devices*, vol. 45, no. 2, pp. 430–438, 1998.

[29] S. S. O. I. Mosfet, "Substrate-Bias Effect and Source – Drain Breakdown Characteristics in Body-Tied," vol. 46, no. 1, pp. 151–158, 1999.

[30] A. Hokazono, S. Balasubramanian, K. Ishimaru, H. Ishiuchi, T. J. K. Liu, and C. Hu, "MOSFET design for forward body biasing scheme," *IEEE Electron Device Lett.*, vol. 27, no. 5, pp. 387–389, 2006.

[31] F. Bias *et al.*, "MOSFET Hot-Carrier Reliability Improvement," vol. 27, no. 7, pp. 605–608, 2006.

[32] T. Rudenko, V. Kilchytska, N. Collaert, M. Jurczak, A. Nazarov, and D. Flandre, "Substrate bias effect linked to parasitic series resistance in multiple-gate SOI MOSFETs," *IEEE Electron Device Lett.*, vol. 28, no. 9, pp. 834–836, 2007.

[33] A. Hokazono, S. Balasubramanian, K. Ishimaru, H. Ishiuchi, C. Hu, and T. J. K. Liu, "Forward body biasing as a bulk-Si CMOS technology scaling strategy," *IEEE Trans. Electron Devices*, vol. 55, no. 10, pp. 2657–2664, 2008.

[34] L. Atzeni and S. Manzini, "Effect of body bias on NBTI of p-MOSFETs," *IEEE Trans. Device Mater. Reliab.*, vol. 17, no. 2, pp. 399–403, 2017.

[35] H. Chen, D. Xie, and L. Guo, "Manipulation of Interface Trap-Induced Generation Current by Substrate Bias in MOSFET," *IEEE Electron Device Lett.*, vol. 39, no. 8, pp. 1126–1128, 2018.

[36] I. Journal and C. Technology, "Reduction of leakage power in CMOS circuits using efficient variable body biasing with bypass technique," vol. 10, no. 38, pp.

4783–4786, 2020.

[37] Z. Tang *et al.*, "Impacts of back gate bias stressing on device characteristics for extremely thin SoI (ETSoI) MOSFETs," *IEEE Electron Device Lett.*, vol. 35, no. 3, pp. 303–305, 2014.

[38] E. G. Marin, F. G. Ruiz, A. Godoy, I. M. Tienda-Luna, C. Martinez-Blanque, and F. Gamiz, "Impact of the back-gate biasing on trigate MOSFET electron mobility," *IEEE Trans. Electron Devices*, vol. 62, no. 1, pp. 224–227, 2015.

[39] O. M. Kane, L. Lucci, P. Scheiblin, T. Poiroux, J. C. Barbe, and F. Danneville, "Back gate impact on the noise performances of 22FDX fully-depleted SOI CMOS," *EuMIC 2020 - 2020 15th Eur. Microw. Integr. Circuits Conf.*, no. January, pp. 81–84, 2021.

[40] A. Gokirmak, "Ultra Narrow Silicon FETS Integrated with Microfluidic System for Serial Sequencing of Biomolecules Based on Local Charge Sensing," no. January, p. 58, 2006.

[41] A. Gokirmak, "Accumulated Body MOSFET," in *Device Research Conference - Conference Digest, DRC*, 2006, pp. 77–78.

[42] A. Gokirmak and S. Tiwari, "Threshold voltage tuning and suppression of edge effects in narrow channel MOSFETs using surrounding buried side-gate," *Electron. Lett.*, vol. 41, no. 3, pp. 157–158, 2005.

[43] M. B. Akbulut *et al.*, "Narrow-channel accumulated-body bulk Si MOSFETs with wide-range dynamic threshold voltage tuning," in *Device Research Conference - Conference Digest, DRC*, 2013, no. SUPPL., pp. 1–2.

[44] M. B. Akbulut *et al.*, "Nanoscale Accumulated Body Si nMOSFETs," *IEEE Trans. Electron Devices*, vol. 65, no. 4, pp. 1283–1289, 2018.

[45] A. Gokirmak and S. Tiwari, "Accumulated body ultranarrow channel silicon transistor with extreme threshold voltage tunability," *Appl. Phys. Lett.*, vol. 91, no. 24, 2007.

[46] Y. Guo *et al.*, "Charge trapping at the MoS2-SiO2 interface and its effects on the characteristics of MoS2 metal-oxide-semiconductor field effect transistors," *Appl. Phys. Lett.*, vol. 106, no. 10, Mar. 2015.

[47] W. F. Clark, B. El-Kareh, R. G. Pires, S. L. Titcomb, and R. L. Anderson, "Low Temperature CMOS—A Brief Review," *IEEE Trans. Components, Hybrids, Manuf. Technol.*, vol. 15, no. 3, pp. 397–404, 1992.

[48] G. Ghibaudo and F. Balestra, "Low temperature characterization of silicon CMOS devices," *Microelectron. Reliab.*, vol. 37, no. 9, pp. 1353–1366, 1997.

[49] C. Luo, Z. Li, T. T. Lu, J. Xu, and G. P. Guo, "MOSFET characterization and modeling at cryogenic temperatures," *Cryogenics (Guildf).*, vol. 98, no. January, pp. 12–17, 2019.

[50] T. Shi, J. Chang, L. Leverich, M. Mallinger, and C. Leader, "1KW push-pull high efficiency RF BJT transistor for radar applications," *IEEE MTT-S Int. Microw. Symp. Dig.*, pp. 1593–1596, 2007.

[51] T. Nomura, M. Masuda, N. Ikeda, and S. Yoshida, "Switching characteristics of GaN HFETs in a half bridge package for high temperature applications," *IEEE Trans. Power Electron.*, vol. 23, no. 2, pp. 692–697, Mar. 2008.

[52] H. Li, C. Yao, C. Han, J. A. Brothers, X. Zhang, and J. Wang, "Evaluation of 600 v GaN based gate injection transistors for high temperature and high efficiency applications," *WiPDA 2015 - 3rd IEEE Work. Wide Bandgap Power Devices Appl.*, pp. 85–91, 2015.

[53] Y. Wu, C. Herrera, A. Hardy, M. Muehle, T. Zimmermann, and T. A. Grotjohn, "Diamond Metal-Semiconductor Field Effect Transistor for High Temperature Applications," *Device Res. Conf. - Conf. Dig. DRC*, vol. 2019-June, no. 2014, pp. 155–156, 2019.

[54] R. M. Incandela, L. Song, H. A. R. Homulle, F. Sebastiano, E. Charbon, and A. Vladimirescu, "Nanometer CMOS characterization and compact modeling at deep-cryogenic temperatures," *Eur. Solid-State Device Res. Conf.*, no. 10, pp. 58–61, 2017.

[55] M. Schwarz, L. E. Calvet, J. P. Snyder, T. Krauss, U. Schwalke, and A. Kloes, "On the Physical Behavior of Cryogenic IV and III-V Schottky Barrier MOSFET Devices," *IEEE Trans. Electron Devices*, vol. 64, no. 9, pp. 3808–3815, 2017.

[56] R. M. Incandela, L. Song, H. Homulle, E. Charbon, A. Vladimirescu, and F. Sebastiano, "Characterization and Compact Modeling of Nanometer CMOS Transistors at Deep-Cryogenic Temperatures," *IEEE J. Electron Devices Soc.*, vol. 6, no. 10, pp. 996–1006, 2018.

[57] A. Beckers, F. Jazaeri, and C. Enz, "Cryogenic MOS Transistor Model," *IEEE Trans. Electron Devices*, vol. 65, no. 9, pp. 3617–3625, 2018.

[58] Z. Wang *et al.*, "Temperature-Driven Gate Geometry Effects in Nanoscale Cryogenic MOSFETs," *IEEE Electron Device Lett.*, vol. 41, no. 5, pp. 661–664, 2020.

[59] P. Chaparro, J. González, G. Magklis, Q. Cai, and A. González, "Understanding the thermal implications of multi-core architectures," *IEEE Trans. Parallel Distrib. Syst.*, vol. 18, no. 8, pp. 1055–1065, Aug. 2007.

[60] M. Prakash Gupta, M. Cho, S. Mukhopadhyay, and S. Kumar, "Thermal investigation into power multiplexing for homogeneous many-core processors," *J. Heat Transfer*, vol. 134, no. 6, 2012.

[61] Y. Lit, B. Lee, D. Brooks, Z. Hu, and K. Skadron, "Impact of thermal constraints on multi-core architectures," *Thermomechanical Phenom. Electron. Syst. -Proceedings Intersoc. Conf.*, vol. 2006, pp. 132–139, 2006.

[62] M. B. Akbulut, "Narrow Channel Accumulated Body MOSFETs : Design , Modeling and Experimental Verification," 2015.

[63] Y. Taur and T. H. Ning, *Fundamentals of Modern VLSI Devices*. Cambridge University Press, 2009.

[64] M. B. Akbulut, H. Silva, and A. Gokirmak, "Three-Dimensional Computational Analysis of Accumulated Body MOSFETs," *IEEE Trans. Nanotechnol.*, vol. 14, no. 5, pp. 847–853, 2015.

[65] D. J. Frank, "Power-constrained CMOS scaling limits," *IBM J. Res. Dev.*, vol. 46, no. 2–3, pp. 235–244, 2002.